\documentclass{mem1}
\usepackage{natbib}\usepackage{txfonts}\usepackage{balance}
\usepackage{graphicx}
\usepackage[a4paper]{hyperref}
\idline{75}{282}
\begin{document}
\def\teff{$T\rm_{eff }$}
\def\kms{$\mathrm {km s}^{-1}$}
\def\lsim{\,\lower2truept\hbox{${<\atop\hbox{\raise4truept\hbox{$\sim$}}}$}\,}
\def\gsim{\,\lower2truept\hbox{${> \atop\hbox{\raise4truept\hbox{$\sim$}}}$}\,}
\def\simlt{\mathrel{\rlap{\lower 3pt\hbox{$\sim$}}\raise 2.0pt\hbox{$<$}}}
\def\simgt{\mathrel{\rlap{\lower 3pt\hbox{$\sim$}} \raise
2.0pt\hbox{$>$}}}
\title{A Physical Model for Co-evolution of QSOs and of their Spheroidal
Hosts  }

   \subtitle{}

\author{G. De Zotti\inst{1,2}
\and F. Shankar\inst{2} \and A. Lapi\inst{2} \and G.L.
Granato\inst{1,2} \and L. Silva\inst{3} \and M. Cirasuolo\inst{2}
\and P. Salucci\inst{2} \and L. Danese\inst{2}
          }

  \offprints{G. De Zotti}

\institute{ INAF -- Osservatorio Astronomico di Padova, Vicolo
dell'Osservatorio 5, I-35122 Padova, Italy
\email{dezotti,granato@pd.astro.it} \and International School for
Advanced Studies, SISSA/ISAS, Via Beirut 2-4, I-34014 Trieste,
Italy \email{lapi,danese,salucci@sissa.it} \and INAF --
Osservatorio Astronomico di Trieste, Via Tiepolo 11, I-34131
Trieste, Italy \email{silva@ts.astro.it }}

\authorrunning{De Zotti et al.}

\titlerunning{Co-evolution of QSOs and spheroidal galaxies}

\abstract{At variance with most semi-analytic models, in the
Anti-hierarchical Baryon Collapse scenario (Granato et al. 2001,
2004) the main driver of the galaxy formation and evolution is not
the merging sequence but are baryon processes. This approach
emphasizes, still in the framework of the hierarchical clustering
paradigm for dark matter halos, feedback processes from supernova
explosions and from active nuclei, that tie together star
formation in spheroidal galaxies and the growth of black holes at
their centers. We review some recent results showing the
remarkably successful predictive power of this scenario, which
allows us to account for the evolution with cosmic time of a broad
variety of properties of galaxies and active nuclei, which proved
to be very challenging for competing models.

 } \maketitle{}

\section{Introduction}

The establishment in the 1990's of the hierarchical clustering
paradigm led to the development of various semi-analytic models
for galaxy formation sharing the basic assumption that the main
driver shaping the structure and morphology of galaxies is
gravity. In this scenario, the gas cools and form stars following
the collapse of dark matter halos. In a cold dark matter
cosmology, small objects form first and merge together to make
larger ones. This scenario then implies that large ellipticals
form late, by the merger of disk/bulge systems made primarily of
stars.

On the other hand, it has long been known that stellar populations
in elliptical galaxies are old and essentially coeval (Sandage \&
Visvanathan 1978; Bernardi et al 1998; Trager et al. 2000;
Terlevich \& Forbes 2002). A color-magnitude relation is also well
established: brighter spheroids are redder (Bower et al. 1992).
The widely accepted interpretation is that brighter objects are
richer in metals and the spread of their star formation epochs is
small enough to avoid smearing of their colors. The slope of this
relation does not change with redshift (Ellis et al. 1997; Kodama
et al. 1998) supporting this interpretation. The star formation
history of spheroidal galaxies is mirrored in the Fundamental
Plane (Djorgovski \& Davies 1987; Dressler et al. 1987) and in its
evolution with redshift.  Elliptical galaxies adhere to this plane
with a surprisingly low orthogonal scatter ($\sim 15$\%), as
expected for a homogeneous family of galaxies. Recent studies
(e.g. Treu et al. 2002; van der Wel et al. 2004; Holden et al.
2004, 2005) suggest that ellipticals, both in the field and in
clusters, follow this fundamental relation up to $z \gsim 1$,
consistent with the hypothesis that massive spheroids are old and
quiescent.

Although these data have mostly to do with ages of stars and may
leave open the issue of the epoch at which ellipticals were
assembled to their present size and mass, they motivated our group
to take a different view, i.e. to investigate the possibility that
gas processes have a key role in driving the formation and
evolution of large galaxies, still in the framework of
hierarchical clustering of dark matter halos (Monaco et al. 2000;
Granato et al. 2001). This new approach emphasized the role of
feedbacks, first of all from supernovae, which release large
amounts of mechanical energy, capable of unbinding the gas in
weakly bound, low mass systems, but also from the active nuclei
which were found to be ubiquitous in the centers of spheroidal
galaxies (Kormendy \& Richstone 1995; Magorrian et al. 1998).
These feedbacks can actually reverse the order of formation of
visible galaxies compared to that of dark halos: large galaxies
form their stars first, while the star formation is delayed in
smaller halos (Anti-hierarchical Baryon Collapse, or ABC,
scenario; Granato et al. 2001, 2004). At the same time, processes
associated to the star formation activity have a profound effect
on the evolutionary history of nuclear activity.

In this paper we will briefly review some recent results obtained
in the framework of the ABC scenario, focussing in particular on
predictions for the evolution of nuclear activity. We adopt a
spatially flat cold dark matter cosmology with cosmological
constant, consistent with the Wilkinson Microwave Anisot\-ro\-py
Probe (WMAP) data (Bennett et al. 2003): $\Omega_m=0.3$,
$\Omega_b=0.047$, and $\Omega_\Lambda=0.70$, $H_0=70{\,\rm
km~s^{-1}~Mpc^{-1}}$, $\sigma_8=0.84$, and an index $n=1.0$ for
the power spectrum of primordial density fluctuations.

\section{Correlations between SMBHs and their
host galaxies}

In the last several years various empirical relationships between
super-massive black hole (SMBH) masses and different properties of
their host galaxies have been derived (but see Novak et al. 2006).
These include correlations with the stellar velocity dispersion
(Ferrarese \& Merritt 2000; Gebhardt et al. 2000; Tremaine et al.
2002; Ferrarese \& Ford 2005), with the mass in stars (H\"aring \&
Rix 2004), with the bulge luminosity (McLure \& Dunlop 2002;
Marconi \& Hunt 2003; Bettoni et al. 2003), with light
concentration (Graham et al. 2001), and with the dark halo mass
(Ferrarese 2002; Baes et al. 2003). In all cases, BH masses are
related to the (generally old) bulge stellar population, not to
the younger disk (Kormendy \& Gebhardt 2001; Kormendy \& Ho 2000;
Salucci et al. 2000).

Shankar et al. (2006) exploited updated estimates of the total and
baryonic mass functions of galaxies, of their luminosity function,
of their velocity dispersion function, and of the mass function of
central black holes to derive relationships between halo masses
and stellar masses, stellar velocity dispersions, and black hole
masses ($M_{\rm BH}$). Particularly relevant, in the present
context are the relationship between $M_h$ and
$M_{\mathrm{star}}$, both in solar units:
\begin{equation}\label{eq|MsMh}
{M_{\mathrm{star}}\over 10^{10}\, M_{\odot}} \approx 2.3 \frac {
(M_h/3\times 10^{11} )^{3.1} }{1+ (M_h/3\times 10^{11})^{2.2}} ~.
\end{equation}
and between $M_{\rm BH}$ and $M_h$:
\begin{equation}\label{eq|MbhMh}
{M_{\rm BH}\over 10^{7} \, M_{\odot}} \approx 0.6\frac
{(M_h/2.2\times 10^{11})^{3.95} }{1+ (M_h/2.2\times
10^{11})^{2.7}}~.
\end{equation}
%

\section{Velocity Dispersion Function and Virial Velocity
Function}\label{sec_VDF}

As first pointed out by Loeb \& Peebles (2003), a comparison of
the local Velocity Dispersion Function (VDF) with the Virial
Velocity Function (VVF) can provide interesting hints on the
structure formation process. Accurate determinations of the VDF of
early-type galaxies have been obtained by Sheth et al. (2003) and
Shankar et al. (2004), based on a large sample ($\sim 9000$ E/S0
galaxies) drawn from the SDSS (Bernardi et al. 2003). As for dark
matter halos, it is convenient to define a 'virial' velocity,
equal to the circular velocity at the virial radius (Navarro et
al. 1997, NFW; Bullock et al. 2001):
\begin{equation}\label{eq_vvir}
V_{\rm vir}^2=\frac{GM_{\rm vir}}{R_{\rm vir}}\ .
\end{equation}
Since, given the virialization redshift, $V_{\rm vir}$ depends
only on $M_{\rm vir}$ ($V_{\rm vir}\propto M_{\rm vir}^{1/3}$) the
VVF can be straightforwardly derived from the mass distribution
function of spheroidal galaxies, integrated over the virialization
redshifts.

Cirasuolo et al. (2005), assuming that all massive halos
($2.5\times 10^{11} M_\odot \simlt M_{\rm vir} \simlt 2\times
10^{13} M_\odot$) virializing at $z \ge 1.5$ yield spheroidal
galaxies or bulges of later type galaxies, showed that the VVF
accurately matches the VDF for a constant ratio of the velocity
dispersion $\sigma$ to the virial velocity, $\sigma/V_{\rm vir}
\simeq 0.55$, close to the value expected at virialization if it
typically occurred at $z\gsim 3$. This is a remarkable result,
since the VVF depends only on the evolution of dark matter halos,
while the VDF is affected by the physics of baryons.

A substantial stability of the halo circular velocity after the
fast accretion phase was found by Zhao et al. (2003) in
high-resolution N-body simulations, even though the halo mass
increases by a substantial factor. But the central velocity
dispersion may be affected by merging events and dissipative
baryon loading.  This strongly suggests that dissipative processes
and later merging events had little impact on the matter density
profile, consistent with the dynamical attractor hypothesis (Loeb
\& Peebles 2003; Gao et al. 2004). It may also suggest that, for
massive galaxies, the mass assembly is largely complete at
substantial redshifts, as also indicated by the {\it Spitzer Space
Telescope} observations by Papovich et al. (2006). Direct evidence
that the galaxy stellar mass function does not evolve
significantly since $z\simeq 1.2$ has been presented by Bundy et
al. (2006).

These results may be relevant also for the interpretation of the
$M_{\rm BH}$--$\sigma$ correlation, which probably originated at
substantial redshifts, when the old stellar population formed. If
so, both $M_{\rm BH}$ and $\sigma$ should not have changed much
since then.

\section{Black hole growth}\label{sec_BHG}

Another important ingredient of the evolutionary history of
quasars is how the active nuclei acquired their mass. Although
radiative accretion must play a role (Soltan 1982), it is not
necessarily the main mechanism for the BH mass growth. In the
hierarchical assembly scenario for massive galaxies a buildup of
SMBHs by coalescence of BHs associated to merging subunits is
naturally expected.

On the other hand, the analysis by Shankar et al. (2004; see also
Marconi et al. 2004), exploiting up-to-date luminosity functions
of hard X-ray and optically selected AGNs, has shown that the
local SMBH mass function is fully accounted for by mass accreted
by X-ray selected AGNs, with bolometric corrections indicated by
current observations and a standard mass-to-light conversion
efficiency $\epsilon \simeq 10\%$. An unlikely fine tuning of the
parameters would be required to account for the local SMBH mass
function accomodating a dominant contribution from `dark' BH
growth (due, e.g., to BH coalescence). It may be noted that the
local SMBH mass function can be rather accurately assessed.
Estimates from either the velocity dispersion function of galaxies
coupled with the $M_{\rm BH}$--$\sigma$ relation or from the
galaxy luminosity function coupled with the $M_{\rm BH}$--$L_{\rm
sph}$ relation of McLure \& Dunlop (2002) agree very well (cf.
Fig.~11 of Shankar et al. 2004).

The visibility time, during which AGNs are luminous enough to be
detected by the currently available X-ray surveys, is found to be
of 0.1--0.3 Gyr for present day BH masses $M_{\rm BH}^0\simeq
10^6$ -- $10^9\,M_{\odot}$.

\section{The Granato et al. (2004) model} \label{sec:gds04}

\subsection{Overview of the model}

While referring to the Granato et al. (2004) paper for a full
account of the model assumptions and their physical justification,
we provide here, for the reader's convenience, a brief summary of
its main features.

The model follows with simple, physically grounded recipes and a
semi-analytic technique the evolution of the baryonic component of
proto-spheroidal galaxies within massive dark matter (DM) halos
forming at the rate predicted by the standard hierarchical
clustering scenario within a $\Lambda$CDM cosmology. The main
novelty with respect to other semi-analytic models is the central
role attributed to the mutual feedback between star formation and
growth of a super massive black hole (SMBH) in the galaxy center.

The idea that SN and  QSO feedback play an important role in the
evolution of spheroidal galaxies has been pointed out by several
authors (Dekel \& Silk 1986; White \& Frenk 1991; Haehnelt,
Natarajan \& Rees 1998; Silk \& Rees 1998; Fabian 1999). Granato
et al. (2004) worked out, for the first time, the symbiotic
evolution of the host galaxy and the central black-hole (BH),
including the feedback. In this model, the formation rate of
massive halos ($2.5\times 10^{11} M_\odot \lsim M_{\rm vir} \lsim
2\times 10^{13} M_\odot$) is approximated by the positive part of
the time derivative of the halo mass function (Press \& Schechter
1974, revised by Sheth \& Tormen 2002). The gas, heated at virial
temperature and moderately clumpy, cools to form stars, especially
in the innermost regions where the density is the highest. The
radiation drag due to starlight acts on the cold gas, decreasing
its angular momentum and causing an inflow into a reservoir around
the central BH, to be subsequently accreted into it, increasing
its mass and powering the nuclear activity. In turn, the feedbacks
from SN explosions and from the active nucleus regulate the star
formation rate and the gas inflow, and eventually unbind the
residual gas, thus halting both the star formation and the BH
growth. Important parameters are the efficiency of SN energy
transfer to the cold gas and the fraction of the QSO luminosity in
winds.

The model prescriptions are assumed to apply to DM halos
virializing at $z_{\rm vir} \gsim 1.5$ and $M_{\rm vir} \gsim 2.5
\times 10^{11} M_\odot$. These cuts are meant to crudely single
out galactic halos associated with spheroidal galaxies. Disk (and
irregular) galaxies are envisaged as associated primarily to halos
virializing at $z_{\rm vir} \lsim 1.5$, some of which have
incorporated most halos less massive than $2.5 \times
10^{11}\,M_\odot$ virializing at earlier times, that may become
the bulges of late type galaxies.

The kinetic energy fed by supernovae is increasingly effective,
with decreasing halo mass, in slowing down (and eventually
halting) both the star formation and the gas accretion onto the
central black hole. On the contrary, star formation and black hole
growth proceed very effectively in the more massive halos, giving
rise to the bright SCUBA phase, until the energy injected by the
active nucleus in the surrounding interstellar gas unbinds it,
thus halting both the star formation and the black hole growth
(and establishing the observed relationship between black hole
mass and stellar velocity dispersion or halo mass). Not only the
black hole growth is faster in more massive halos, but also the
feedback of the active nucleus on the interstellar medium is
stronger, to the effect of sweeping out such medium earlier, thus
causing a shorter duration of the active star-formation phase.

The basic yields of the model are the star-formation rate,
$\psi(t)$, as a function of the galactic age, $t$, (hence the
evolution of the mass in stars, $M_{\rm sph}(t)$), and the growth
of the central BH mass, $M_{\rm BH}(t)$, for any given value of
the halo mass, $M_{\rm vir}$, and of the virialization redshift,
$z_{\rm vir}$. These quantities are obtained solving the system of
differential equations given by Granato et al. (2004) and
Cirasuolo et al. (2005).

\begin{figure}[t!]
\resizebox{\hsize}{!}{\includegraphics[clip=true]{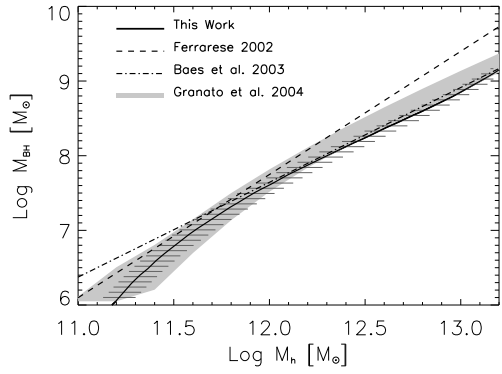}}
\caption{\footnotesize Comparison of the relationship between
$M_{\rm BH}$ and $M_{\rm halo}$ derived by Shankar et al. (2006,
solid line; the thin horizontal bars show the uncertainties) with
the prediction of the model by Granato et al. (2004; shaded area)
and the observation-based estimates by Ferrarese (2002; dashed
line) and Baes et al. (2003; dot-dashed line), referred to a
virialization redshift $z_{\rm vir}=3$.  } \label{Mbh_Mhalo_zvir}
\end{figure}

\subsection{Approximate analytic solutions}

A simplified set of equations, still keeping the basic physics, is
presented by Shankar et al. (2006), who derived an approximate
analytic solution for the time evolution of the mass in stars,
$M_{\mathrm{star}}$, which, at the present time $t_0$, writes:
\begin{equation}\label{eq|Ms}
M_{\mathrm{star}}(t_0)=f_{\mathrm{surv}}\frac{f_{\mathrm{cosm}}\,M_h(0)}{\gamma},
\end{equation}
where $f_{\mathrm{surv}}$ is the fraction of stars still surviving
at the present time ($f_{\mathrm{surv}}\approx 0.6$ for a Salpeter
IMF after about 10 Gyr from a burst), $f_{\mathrm{cosm}}\,M_h(0)$
is the initial gas mass  ($f_{\mathrm{cosm}}\simeq 0.17$ being the
cosmic baryon to dark matter ratio and $M_h$ being the halo mass).
Further, $\gamma=1-R+\alpha$, $R$ being the fraction of mass
restituted by evolved stars ($R\approx 0.3$ for a Salpeter IMF),
and
\begin{equation}\label{eq|SNfeed}
\alpha=\frac {N_{\rm SN}\, \epsilon_{\rm SN}\,E_{\rm SN}}{E_B}
\end{equation}
is effective efficiency for the removal of cold gas by the
supernova feedback. Here $N_{\rm SN}$ is the number of SNe per
unit solar mass of condensed stars, $\epsilon_{\rm SN}\, E_{\rm
SN}$ the energy per SN used to remove the cold gas, and $E_B$ the
binding energy of the gas within the DM halo, per unit gas mass.

For $z\gsim 1$ Shankar et al. (2006) obtain:
\begin{eqnarray}\label{eq|SNeff}
\alpha & \approx & 1.2 \, \left({N_{\rm SN}/8\times
10^{-3}}\right)\, \left({\epsilon_{\rm SN}/ 0.1}\right)\,\cdot \nonumber \\
&\cdot& \left({E_{\rm SN}/ 10^{51}\hbox{erg}}\right)\,
 \left[(1+z)/ 4\right]^{-1}\,\cdot \nonumber \\ &\cdot& \left({M_h/ 10^{12}\,
M_{\odot}}\right)^{-2/3}.
\end{eqnarray}
For large halo masses, where the stellar feedback is less
efficient ($\alpha \la 1$), the quantity $1-R+\alpha$ is a slowly
decreasing function of the halo mass, so that $M_{\mathrm{star}}$
is approximately proportional to $M_h$. However, in this case, the
fraction of gas turned into stars is controlled by the AGN
feedback, which, as shown by the full treatment by Granato et al.
(2004), for $M_h \ga 3 \times 10^{11}\, M_\odot$ expels an
approximately constant fraction of the initial gas, thus
preserving the approximate proportionality between
$M_{\mathrm{star}}$ and $M_h$, in agreement with
eq.~(\ref{eq|MsMh}).

On the contrary, for $M_h \ll 10^{12}\, M_{\odot}$, $\gamma\simeq
\alpha \propto M_{h}^{-2/3}$, and
\begin{equation}\label{eq|slopeMsMh}
M_{\mathrm{star}}\propto f_{\mathrm{surv}}\, M_h^{5/3},
\end{equation}
quantifying the effect of SN in decreasing the
$M_{\mathrm{star}}/M_h$ ratio with decreasing $M_h$.

In the Granato et al. (2004) model, the radiation drag dissipates
the angular momentum of the cool gas present in the central
regions (Kawakatu \& Umemura 2002). The gas then falls into a
reservoir around the central BH  at a rate
\begin{equation}\label{eq|dotMres}
\dot{M}_{\mathrm{res}}=1.2 \times 10^{-3}\psi(t)(1-e^{-\tau})~,
\end{equation}
where $\tau$ is the effective optical depth of the central star
forming regions. If most of the mass in the reservoir is
ultimately accreted onto the central BH, and if other
contributions to the BH mass can be neglected, we have:
\begin{equation}\label{eq|Mbh}
{M}_{BH}\approx 1.2 \times 10^{-3}\, M_{\mathrm{star}}\,
(1-e^{-\tau})~.
\end{equation}
Granato et al. (2004) assumed that the effective optical depth
depends on the cold gas metallicity and mass $\tau\propto Z\,
M_{\mathrm{gas}}^{1/3}$. The outcome of their numerical code
yields, on average, $Z\propto M_h^{0.3}$ in the mass range
$10^{11}\, M_{\odot}\leq M_h \leq 3 \times 10^{13}\, M_{\odot}$
[cf. their Figs.~(5) and (8)]. Since $M_{\mathrm{gas}}\sim
f_{\mathrm{cosm}} M_h$, one gets
\begin{equation}\label{eq|optdepth}
\tau \propto M_h^{2/3} ~.
\end{equation}
In the mass range $M_h\la  10^{11}\, M_{\odot}$ the optical depth
is small ($\tau \ll 1$); from eqs.~(\ref{eq|Mbh}) and
(\ref{eq|optdepth}) we then obtain:
\begin{equation}\label{eq|slopeMbhMh}
M_{\rm BH}\propto M_{\mathrm{star}}\, \tau \propto M_h^{7/3}~.
\end{equation}
Thus this simple model predicts that the low mass slope of the
$M_{\rm BH}$--$M_h$ relation is steeper than that of the
$M_{\mathrm{star}}-M_h$ relation, because of the decrease of the
optical depth with mass $\tau\propto M_h^{2/3}$, entailing a lower
capability of feeding the reservoir around the BH.

For large halo masses $\tau \gg 1$, and the model predicts $M_{\rm
BH} \propto M_{\mathrm{star}} \propto M_h$. A roughly linear
relationship between $M_{\rm BH}$ and $M_{\mathrm{star}}$ ($M_{\rm
BH} \propto M_{\mathrm{star}}^{1.12\pm 0.06}$) has been
observationally derived by H\"aring \& Rix (2004).

In Fig.~\ref{Mbh_Mhalo_zvir} we compare the $M_{\rm BH}$--$M_h$
relation derived by Shankar et al. (2006) with that of Ferrarese
(2002), who first investigated this issue from an observational
point of view. She derived a power-law relationship between the
bulge velocity dispersion and the circular velocity, $v_c$, for a
sample of spiral and elliptical galaxies spanning the range
$100\la v_c\la 300$ km s$^{-1}$, and combined it with an
approximated relationship between $v_c$ and the virial velocity,
$v_{\rm vir}$, based on the numerical simulations by Bullock et
al. (2001) and with the $M_h$--$v_{\rm vir}$ relationship given by
the $\Lambda$CDM model of the latter authors for a virialization
redshift $z_{\rm vir}\sim 0$, to obtain a $M_h$--$\sigma$
relation. Coupling it with one version of the observed BH mass vs.
stellar velocity dispersion relationship ($M_{\rm BH}\propto
\sigma^{4.58}$) she obtained $M_{\rm BH} \propto M_{h}^{\alpha}$,
with $\alpha= 1.65$--1.82. Baes et al. (2003) with the same
method, but assuming $M_{\rm BH}\propto \sigma^{4.02}$ and with
new velocity dispersion measurements of spiral galaxies with
extended rotation curves, yielding a slightly different
$v_c$-$\sigma$ relation, found $M_{\rm BH} \propto M_{h}^{1.27}$.

As shown by Fig.~6$a$ of Shankar et al. (2006), their
relationships lie substantially below ours.\footnote{Both
Ferrarese (2002) and Baes et al. (2003) adopted
$V_{\mathrm{vir}}\propto
[\Omega_M(\Delta_{\mathrm{vir}}/200)]^{\alpha}$ with $\alpha=1/3$,
following Bullock et al. (2001), while the correct value is
$\alpha=1/2$. This correction has, however, a minor effect.} It
should be noted, however, that the $M_h$--$v_{\rm vir}$ relation
depends on the virialization redshift. For $z_{\rm vir} \simeq 3$,
its coefficient would be a factor of $\simeq 4.25$ lower than that
used by Ferrarese (2002) and Baes et al. (2003) and the
coefficients of the $M_{\rm BH} \propto M_{h}$ relations would be
larger by a factor of $\simeq 5.6$ in the case of eq.~(6) of
Ferrarese(2002) or of $\simeq 4$ in the case of Baes et al.
(2003),  bringing them much closer to the results by Shankar et
al. (2006; see Fig.~\ref{Mbh_Mhalo_zvir}), and to the predictions
of the Granato et al. (2004) model.

\begin{figure*}[t!]
\resizebox{\hsize}{!}{\includegraphics[clip=true]{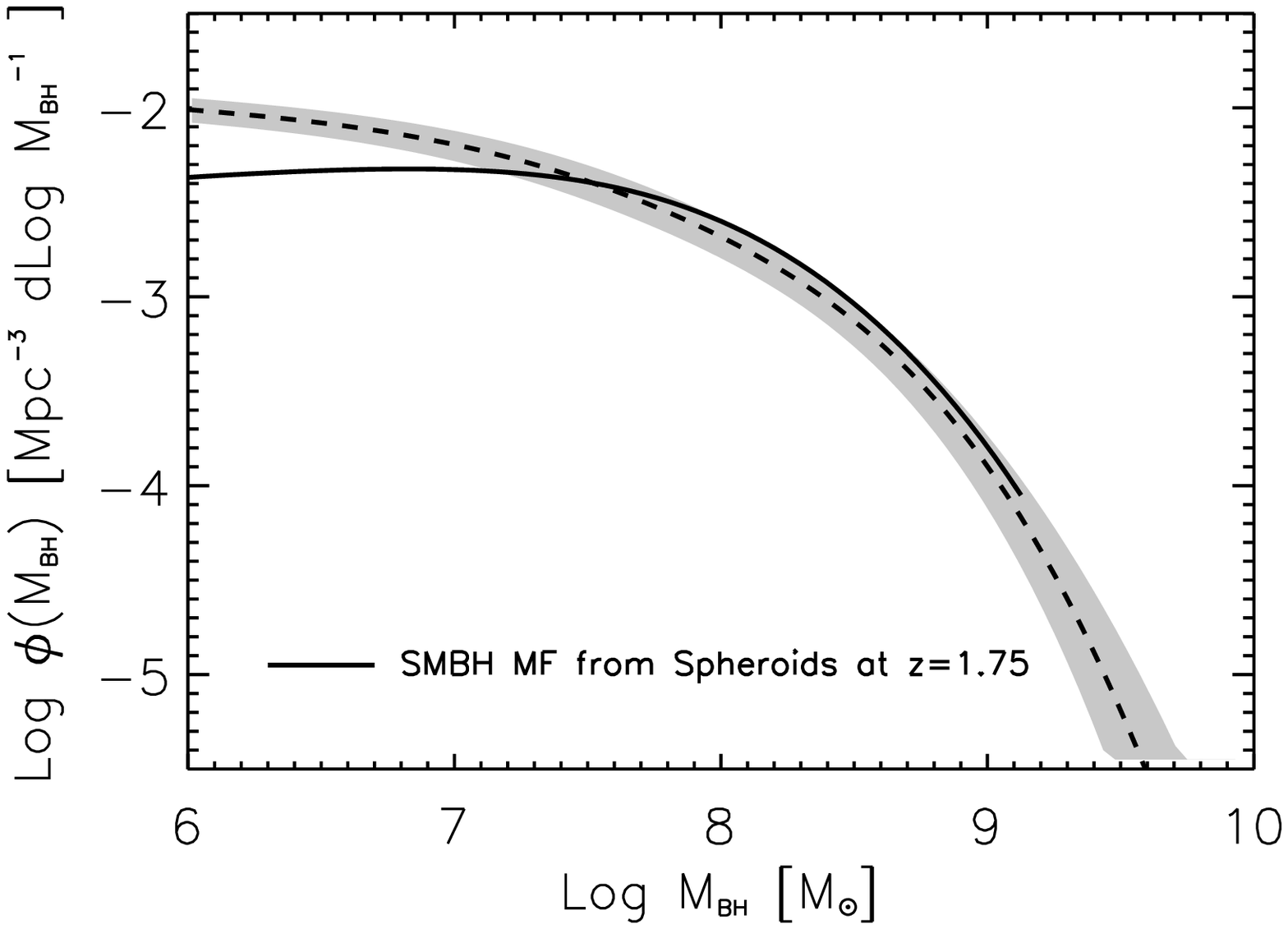}
\includegraphics[clip=true]{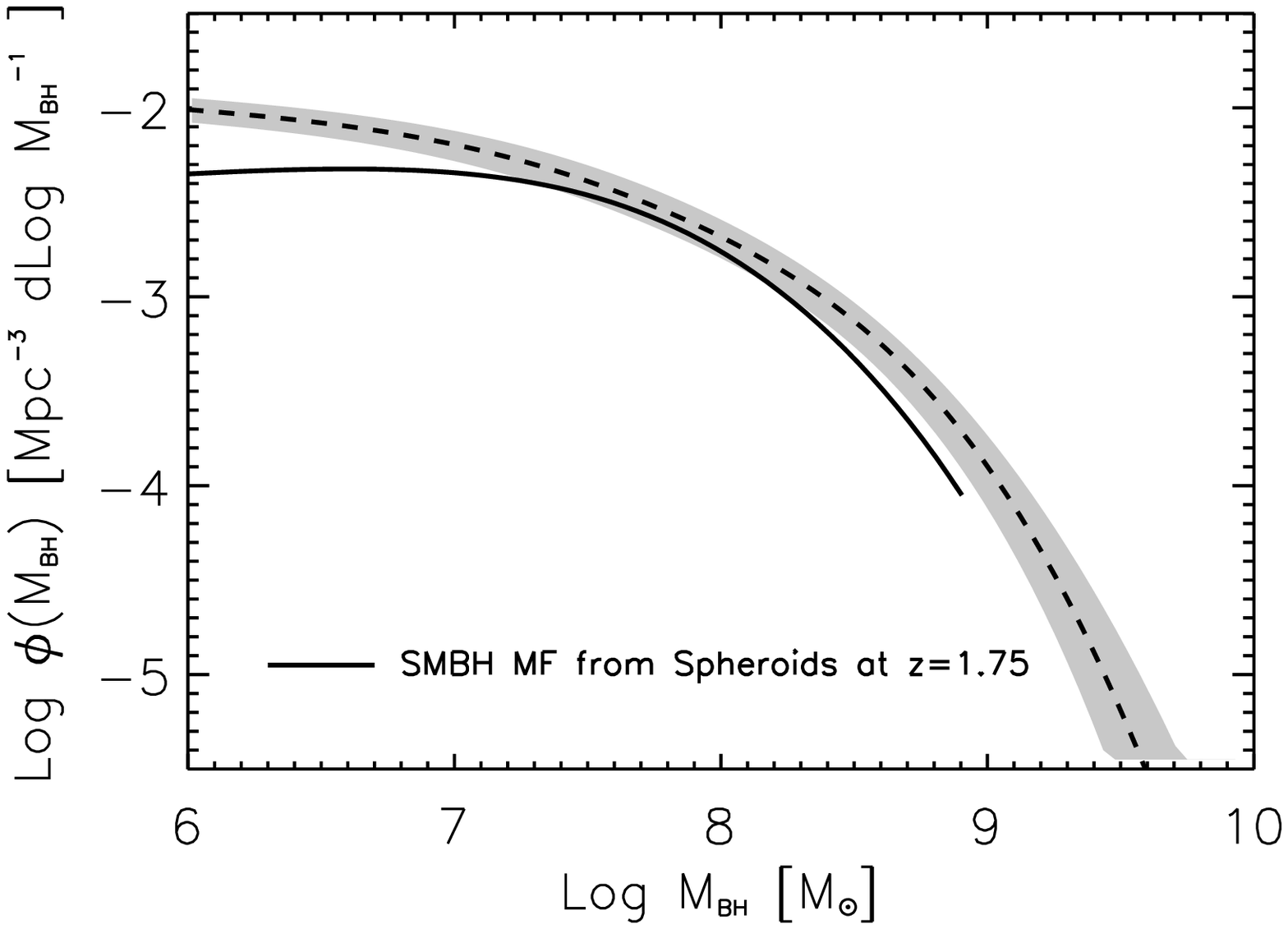}}
\caption{\footnotesize Contribution of SMBH associated to
spheroidal galaxies at $z\simeq 2$ (solid line) to the local SMBH
mass function (dashed line; the shaded area shows the uncertainty)
estimated by Shankar et al. (2004), assuming a $M_{\rm
BH}/M_{\mathrm{star}}$ ratio 3 times higher than in the local
universe and adopting a $M/L$ ratio appropriate for a Salpeter
(left-hand panel) IMF or an IMF flattening below $1\,M_\odot$
(right-hand panel), with a dispersion of 0.3 dex. }
\label{SMBH_MF}
\end{figure*}

\subsection{The evolution of nuclear activity}

According to the Granato et al. (2004) model the SMBH growth
parallels the star formation in spheroidal galaxies. However, the
BH growth is much faster than the increase of the mass in stars.
In fact, the timescale for the former is roughly the $e$-folding
timescale for Eddington-limited accretion, $t_{\rm ef}=t_E\,
\epsilon /(1-\epsilon)$ (where $t_E$ is the Eddington time and
$\epsilon$ is the BH mass to energy conversion efficiency). For
$\epsilon=0.1$, $t_{\rm ef}\simeq 5\times 10^7\,$yr. The star
formation timescale, which is the shorter between the cooling and
the free-fall times, increases from the denser inner regions
outwards (where most of the mass is); its effective values are
typically of a few to several hundred million years. This means
that the $M_{\rm BH}/M_{\mathrm{star}}$ ratio increases rapidly
until the AGN feedback sweeps out the residual gas.

The exponential growth of the active nucleus implies that its
bolometric luminosity is very low except in the final several
$e$-folding times. Also the nucleus is surrounded by a dusty
environment with a large column density, even if we ignore the
contribution of the obscuring torus. For example, if we adopt for
the gas distribution a King profile with core radius $\sim
0.3$--$1\,$kpc, we obtain, for galaxies with $M_{h} \gsim 2 \,
10^{12}\,M_\odot$, column densities to the nucleus of $N_H \gsim
\mbox{a few} \times 10^{23}$ to $\mbox{a few}\times
10^{24}\,\hbox{cm}^{-2}$ (Granato et al. 2006). Correspondingly,
we expect large optical extinctions $A_V \sim 200 (N_H/ 3 \times
10^{23}\,\hbox{cm}^{-2})$.

The nuclear activity then shows up first in hard X-rays. As noted
in \S$\,$\ref{sec_BHG}, the X-ray visibility times are in the
range 0.1--0.3 Gyr, corresponding to the last 2--6 $e$-folding
times of the BH growth. In this period, the $M_{\rm
BH}/M_{\mathrm{star}}$ ratio can be $\ll 1$. The AGN feedback will
then progressively sweep out the residual gas, decreasing the
extinction until the AGN enters its optical visibility phase,
whose duration is expected to be a substantial fraction of the
last $e$-folding time.

These expectations are fully borne out by the study of 20 SCUBA
galaxies brighter than $\simeq 4\,$mJy at $850\,\mu$m, lying in
the region of the 2Ms CDF-N Chandra observations (Alexander et al.
2003, 2005a,b). As shown by Granato et al. (2006), the data imply
accretion rates and column densities fully compatible with the
predictions of the Granato et al. (2004) model. Borys et al.
(2005) estimate, for these objects, $M_{\rm BH}/M_{\mathrm{star}}$
ratios one or two orders of magnitude smaller than local galaxies
of comparable stellar mass, suggesting a subsequent growth lasting
2 to 5 $e$-folding times, at the Eddington limit. The lack of
objects with $M_{\rm BH}/M_{\mathrm{star}}\gsim 0.1 (M_{\rm
BH}/M_{\mathrm{star}})_{\rm local}$ may suggest that in the last
$\sim 2$--3 $e$-folding times the AGN feedback begins quenching
the star formation rate so that the sub-mm emission decreases and
galaxies slip below the $850\,\mu$m flux limit.

A different history of the coevolution of SMBHs and host galaxies
is apparently implied by the study by Peng et al. (2006), who
estimated the $M_{\rm BH}/M_{\mathrm{star}}$ ratios for a sample
of optical quasars at $z \sim 2$ and found them to be 3--6 times
{\it larger} than in the local universe. This conclusion however
relies on a number of very uncertain assumptions, as discussed by
the authors themselves. A further constraint come from the SMBH
local mass function (see \S$\,$\ref{sec_BHG}). An estimate of the
galaxy stellar mass function at $z\simeq 2$ has been obtained by
Drory et al. (2005). Others can be obtained from the galaxy
luminosity functions at a similar redshift derived by various
groups (Saracco et al. 2006; Dahlen et al. 2005; Fontana et al.
2004; Caputi et al. 2005). As shown by Fig.~\ref{SMBH_MF}, a
$M_{\rm BH}/M_{\mathrm{star}}$ ratio 3 times higher than in the
local universe would imply that, at large SMBH masses, there is no
room for growth at $z\lsim 2$. The constraint is attenuated if the
IMF flattens to a slope of $-0.4$ below $1\, M_\odot$ (see Romano
et al. 2002 and Granato et al. 2004), decreasing the $M/L$ ratio
in the $K$-band to $\simeq 0.3$ (this ratio is $\simeq 0.5$ for a
Salpeter IMF). Nevertheless, even in this case a rather large
contribution to the local SMBH mass function is already in place
at $z\simeq 2$, and this is very difficult to reconcile with the
optical and X-ray data on the evolution of the AGN luminosity
function.

\section{Conclusions}

Huge amounts of data have been accumulating in recent years,
providing rather detailed information on the mass-dependent
evolution of galaxies and of active nuclei. Rather unexpectedly,
it turned out that intense star formation shifts from massive
galaxies at high redshifts to lower mass galaxies at later time, a
pattern referred to as ``downsizing'' (Cowie et al. 1996). Granato
et al. (2001, 2004) showed that this evolutionary behaviour, which
apparently contradicts expectations from the hierarchical
structure formation paradigm, can be reconciled with it if the
main driver of evolution is not the merging sequence of dark
matter halos, but are baryon processes, involving strong feedback
effects from star formation (and supernova explosions) and from
nuclear activity (Anti-hierarchical Baryon Collapse - ABC -
scenario).

An implication of this scenario is that the luminosity function of
massive galaxies underwent essentially passive evolution since
$z\simeq 1.5$, little affected even by the ``dry mergers"
advocated by several authors (e.g. Bell et al. 2005; Faber et al.
2006; van Dokkum 2005), but strongly constrained by the DEEP2
Galaxy Redshift Survey (Bundy et al. 2006). The ABC scenario also
explains a broad variety of observational data, as shown by
Granato et al. (2004), Silva et al. (2005), Cirasuolo et al.
(2005), and Shankar et al. (2006). We have recalled here the close
match of the virial velocity function with the velocity dispersion
function of galaxies, and the $M_{\rm BH}$--$\sigma$ relation,
both determined from data on local galaxies, but whose physical
explanations trace their origin to substantial redshifts.

In this framework, also the AGN evolution is tightly connected
with star formation, rather than with mergers (although star
formation may be triggered by merging events), and the BH growth
is dominated by radiative accretion, rather than, e.g., by
coalescence of BH associated to merging sub-units. This implies
that during the intense star-formation phases, characterizing
sub-mm bright galaxies, the BH masses are still below, even by
large factors, the present day values, consistent with the
findings by Alexander et al. (2003, 2005a,b) and Borys et al.
(2005). Also, the detectable QSO activity is {\it delayed}
compared to the onset of vigorous star formation, since: a) the BH
mass and its bolometric luminosity grow exponentially on a
timescale $t_{\rm ef}\simeq 5\times 10^7\,$yr and are therefore
small except in the final few tenths of Gyr before reaching the
final mass; b) the nuclear activity is highly obscured even in
hard X-rays before the AGN feedback begins to push away the
residual dusty medium.

\begin{acknowledgements}
Work supported in part by ASI and MIUR.
\end{acknowledgements}

\bibliographystyle{aa}

\end{document}